\documentclass[preprint, amsmath,amssymb, aps]{revtex4-2}
\usepackage{graphicx}
\usepackage{dcolumn}
\usepackage{bm}
\usepackage{xcolor}
\usepackage{lineno}
\usepackage{ulem}
\newcommand{\avg}[1]{\left\langle #1  \right\rangle}

\begin{document}

\title{Complexity emerges in measures of the marking dynamics in football games}

\author{A. Chacoma}
\email{achacoma@famaf.unc.edu.ar}
\affiliation{Instituto de F\'isica Enrique Gaviola (IFEG-CONICET), Ciudad Universitaria, 5000 Córdoba, Argentina}
\affiliation{Facultad de Matem\'atica, Astronom\'ia, F\'isica y Computaci\'on, Universidad Nacional de C\'ordoba, Ciudad Universitaria, 5000 Córdoba, Argentina}

\author{M. N. Kuperman}
\affiliation{Instituto Balseiro, Universidad Nacional de Cuyo, R8402AGP Bariloche, Argentina}
\affiliation{Centro At\'{o}mico Bariloche and CONICET, R8402AGP Bariloche, Argentina}

\author{O.V. Billoni}
\affiliation{Instituto de F\'isica Enrique Gaviola (IFEG-CONICET), Ciudad Universitaria, 5000 Córdoba, Argentina}
\affiliation{Facultad de Matem\'atica, Astronom\'ia, F\'isica y Computaci\'on, Universidad Nacional de C\'ordoba, Ciudad Universitaria, 5000 Córdoba, Argentina}

\begin{abstract}
In this article, we study the dynamics of marking in football matches. 
To do this, we surveyed and analyzed a database containing the trajectories of players from both teams on the field of play during three professional games. 
We describe the dynamics through the construction of temporal bipartite networks of proximity. 
Based on the introduced concept of proximity, the nodes are the players, and the links are defined between opponents that are close enough to each other at a given moment. 
By studying the evolution of the heterogeneity parameter of the networks during the game, we characterized a scaling law for the average shape of the fluctuations, unveiling the emergence of complexity in the system. Moreover, we proposed a simple model to simulate the players’ motion in the field from where we obtained the evolution of a synthetic proximity network. 
We show that the model captures with a remarkable agreement the complexity of the empirical case, hence it proves to be helpful to elucidate the underlying mechanisms responsible for the observed phenomena.
\end{abstract}

\maketitle
\section{Introduction}

In recent years, the interest in studying the emergence of complexity in team sports competition has increased \cite{ribeiro2012anomalous,clauset2015safe,kiley2016game,ibanez2018relative,ruth2020dodge,chacoma2020modeling,moritz2021risk,yamamoto2021preferential,chacoma2022simple}.
Prompted by the advances in data acquisition and theoretically supported by state-of-the-art statistical tools and artificial intelligence techniques, this area has gone beyond the academy boundaries to positioning as a new vigorous precursor of the innovative processes in the sports industry \cite{rajvsp2020systematic}.

Particularly, in the game of football, the use of network science to describe the dynamics of a match is currently ubiquitous, especially the utilization of the so-called passing networks \cite{gonccalves2017exploring,yamamoto2018examination,ichinose2021robustness}. 
In that framework, the information to set the network's links is given by the number of passes between teammates.
The network structure, in this context, allows analysts to quantify the interaction in the field and the team performance via the use of classical network metrics like the clustering coefficient, the shortest path length, or the eigenvector centrality, among others \cite{martinez2020spatial,buldu2019defining}.

However, this approach considers only the interaction among teammates ignoring the interaction between opponents, i.e., neglecting the effect of the marking dynamic.
In this sense, the use of network science to describe the interaction between opponents has been rarely reported \cite{sugerida1}.
In this context, it is necessary to highlight that the base of the tactical system in the game of football is the marking \cite{sampaio2012measuring}, since it defines the strategy of the team. 
Hence, to carry out a complete analysis of the game it is crucial to characterize this phenomenon.

In this paper, we aim to study the marking dynamics using network science.
To do so,  we survey a database containing the coordinates of the players in the field at each second of three professional games.
With this information, we define a bipartite graph \cite{newman2018networks} where the nodes are the players of both teams, but the connections can be only between opponents. Bipartite networks are a special kind of network where there are two distinct sets of nodes, and connections can only be present between nodes belonging to different categories. There are multiple examples of these networks in an ecological context \cite{dormann09, huaylla21} and several techniques have been developed to generalize the methodologies usually applied in the analysis of monopartite networks \cite{barber07,dormann14,saracco17}.
To establish the connections in our proximity networks, we will use the euclidean distance of the players in the field since the opponents' closeness is strictly related to the marking \cite{low2021porous}.
This particular type of graph is known as {\it proximity network} and has been widely used to study other phenomena in complexity science \cite{isella2011s,farine2015proximity,cattuto2010dynamics,gauvin2013activity,cencetti2021digital}.

The manuscript is divided into three parts: In the next section, we describe the database and give further information on the acquisition process. 
In section results, we first study the evolution of the proximity networks during the game. In our analysis, we observe and characterize statistics regularities that confirm the emergence of complexity in the system.
Secondly, we propose a model to simulate the players' motion in the field and analyze the outcomes by performing the same analysis we used to study the empirical data. Despite the model's simplicity, we achieve a satisfactory performance, obtaining agreement with the observed in the real case. 
In the last section, our main results are briefly summarized.

\section{Data}

We use tracking data from three professional football games provided by the company {\it Metrica Sports} \cite{metrica}. 
They use artificial intelligence applied to visual recognition to gather from video records the players' coordinates in the pitch of both teams  with high resolution.
According to the specification provided on the company website, they work with a resolution of 25 frames per second, 10 cm data range position, and 100 \% identity accuracy.
The data is separated into three datasets, hereafter referred to as $DS1$, $DS2$, and $DS3$, and is publicly available for the community in \cite{data@metrica}. 
All the data is anonymized, there are no references to players' names, teams, or matches. 
In each game, they report 22 tracked players, 11 of one team and 11 of the other including the goalkeepers. The pitch resolution is 105x68 meters.
Notice that the provided temporal resolution is $0.04~sec$. However, we averaged the data to avoid noise, preserving a resolution of $1~sec$.
Thus, the total number of temporal points for $DS1$, $DS2$, and $DS3$ is 5800, 5646, and 5750, respectively.

\section{Results and discussion}

\subsection{Interaction distance}

\begin{figure*}[t!]
\centering
\includegraphics[width=\textwidth]{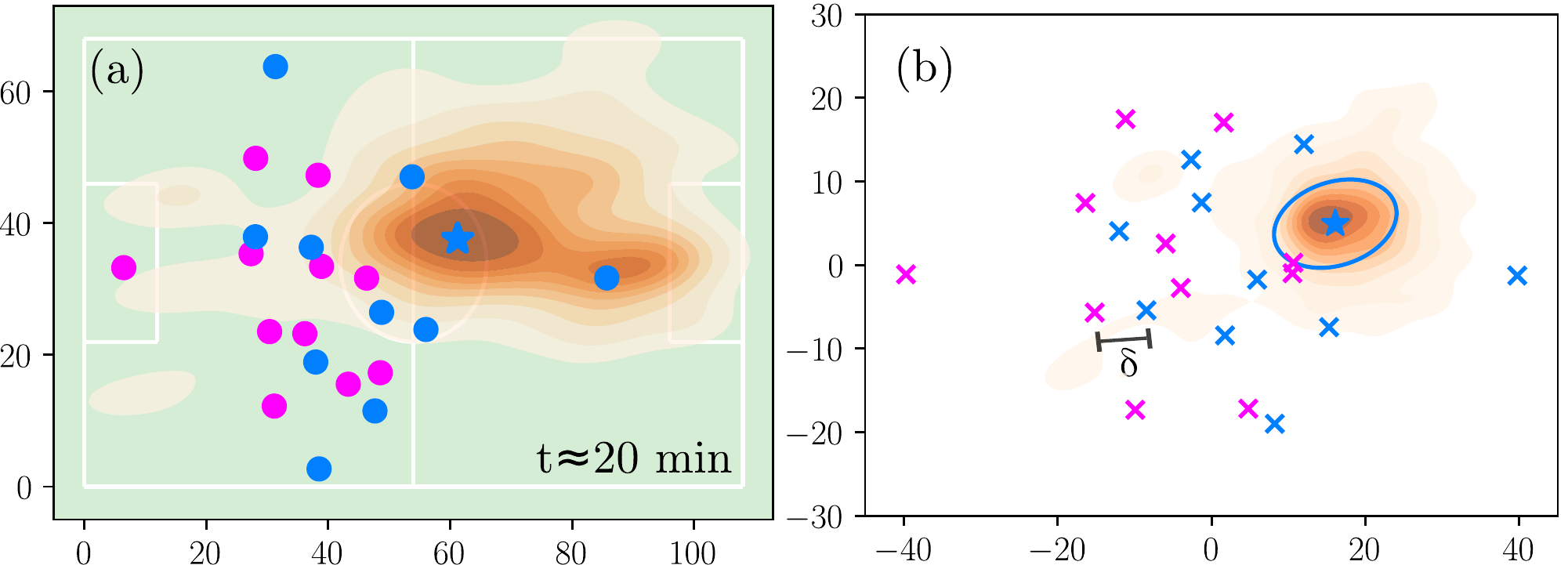}
\caption{
Defining the range of interaction distances.
For this visualization, we use the game recorded in $DS1$.
Panel (a) shows the players position at $t\approx20$. The heatmap in the background shows the explored zones for the player highlighted with a star.
Panel (b) exhibits the average position of the players from the center of mass frame of reference. Parameter $\delta$ indicates the distance between nearest opponents, and the ellipse shows the action zone of the player star.
}
\label{fi:1}
\end{figure*}

In this section, we focus on defining the scope of the interaction range between players. 
With this aim, we describe some aspects of players' movement during the first half of the game recorded in $DS1$.
Let us focus in Fig.~\ref{fi:1}. In panel (a), we show the players on the field at $t\approx20~min$. %
The heat map in the background shows the explored areas by the player of team 1 highlighted with a $star$, giving an example of the typical players' motion in the field.
Since both teams move around in a single block, like a bird flock, it is helpful to analyze the system from the center of mass reference frame.
From this perspective, we calculated the players' average position, hereafter referred to as $\vec{c}_n$, where $n$ indicates the $nth$ player in the field. The result is shown in panel (b).
The ellipse surrounding the player star is calculated from the cloud of points given by all the positions explored by it over time.
The ellipse's center corresponds to the player's center, and the radii $r_1$ and $r_2$ to the standard deviation intervals calculated on the two principal components of the cloud, obtained via a $PCA$ analysis. 
Note that the ellipse measures the range of the player's movement. We will call to this {\it player's action zone}.

We define $\delta$ as the distance between a player's average position and that of its closest opponent. 
Note that this parameter is an emergent of the marking dynamics and gives a measure of the interaction distances between nearest opponents.
Since the marking dynamic is diverse due to the different players involved, we obtain a wide variety of values for $\delta$. However, all the values fall into the small interval $(3.88, 12)~m$, neglecting the contribution of the goalkeepers who do not mark opponents. The same calculation in datasets $DS2$ and $DS3$ gives similar results.
Therefore, we can conclude that this is a good measure for the range of interaction distances where the marking dynamic occurs.

\subsection{Proximity Networks}

We now focus on describing the marking dynamics with a bipartite temporal proximity network.
In this frame, the nodes are the players of both teams, and the links will be only between players of different teams.
To establish a link, we proceed as follows: At every second of the game, we compute the $2D$ euclidean distance between all the pairs of opponents in the field. When the distance between two opponents is lower than a given threshold $\theta$, then we will set a link between them.

\begin{figure*}[t!]
\centering
\includegraphics[width=\textwidth]{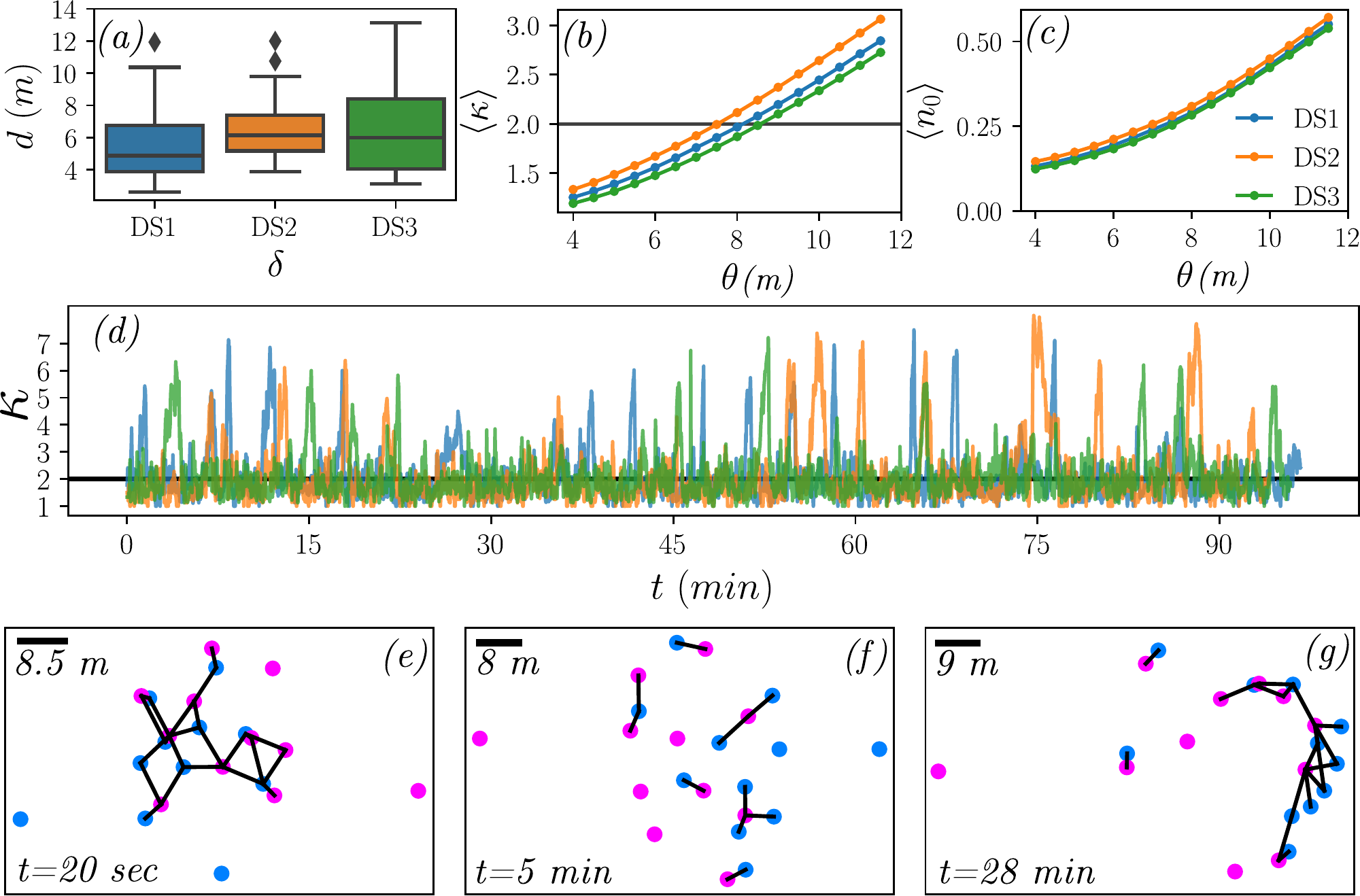}
\caption{
Describing the marking as proximity networks for the three games recorded in $DS1$, $DS2$, and $DS3$.
Panel (a) shows the range of interaction distances used to define the threshold range.
Panel (b) exhibits the relation between the mean value of the heterogeneity parameter, $\avg{\kappa}$, and the threshold $\theta$.
Panel (c) shows the relation between the mean value of the fraction of nodes on the giant component, $\avg{n0}$, and the threshold $\theta$.
Panel (d) the time evolution of the heterogeneity parameter during the game.
In panels (e), (f), and (g), we show the proximity networks displayed in the field a different times for datasets $DS1$, $DS2$, and $DS3$, respectively.
}
\label{fi:2}
\end{figure*}

To select the threshold value in our study, we explore the range of interaction distances defined in the previous section. 
Notice that the three games under analysis have a different range but are still very similar, as we can observe in the box plots of Fig.~\ref{fi:2}~(a).
For every value of $\theta$, we define a network at every second and compute the heterogeneity parameter $\kappa(t)=  \frac{\avg{k^2}}{\avg{k}}$, where $k$ is the node degree. Then we calculate the mean value over time and obtain the relation $\theta$ vs.
$\avg{\kappa}$ showed in Fig.~\ref{fi:2}~(b).
We observe a smooth evolution where, from the range of analyzed values of $\theta$, we obtain values of $\avg{\kappa} $ into the interval $(1,3)$.
The black horizontal line in the plot shows the theoretical percolation point,  $\kappa=2$, derived by Molloy and Reed in \cite{molloy1995critical}.
For networks with no degree correlation in the thermodynamic limit, this point defines a continuous phase transition, where, for $\kappa <2$, all the components in the network are small clusters of trees, and for $\kappa >2$, a giant component of size proportional to $N$ emerges.
In our research, since the networks are relatively small, we cannot frame our results in the theory of phase transitions. 
However, this result evidences the emergence of complexity in the system.
Additionally, Fig.~\ref{fi:2}~(c), aiming to illustrate the transition from small clusters to a large ones, we show the mean value of the fraction of nodes in the giant component, $\avg{n_0}$, as a function of $\theta$.
We can see that for $\theta\approx4~m$, on average less than $20\%$ of the nodes are in giant component, whereas for $\theta\approx12~m$, more than $50\%$.

We now turn to analyze the temporal evolution of the heterogeneity parameter $\kappa$.
Since we aim to compare the curves of the three datasets, we tune the value of $\theta$ in each case such that $\avg{\kappa}\approx 2$.
With this approach we obtain the values $\theta_1=8.5$, $\theta_2=8$ and $\theta_3=9$.
In Fig.~\ref{fi:2}~(d), we show the evolution of $\kappa$ during the $90$ minutes (plus extra time) of the three games. As expected, we can observe that the values fluctuate around $\kappa=2$. This behavior indicates that the network structure oscillates between periods of high clusterization and high defragmentation.
In the supplementary material, we incorporate animation to visualize this result.
The higher peaks in the series correspond to special situations of the game where the players group altogether, like, for instance, in a corner kick or dead ball \cite{pulling2013defending}.
In this regard, we want to highlight that $\kappa$ cannot grow to the infinite. There is a limit graph given by all the opponents connected with 22 nodes, 121 edges, and $\kappa=11$.
We can also define a static graph by using the average position of the players. For instance, for $DS1$, we have a static graph with 17 nodes, 16 edges, and $\kappa=2.25$. Note $\kappa \approx 2$, as expected.

Finally, in Fig.~\ref{fi:2} panels (e), (f), and (g), we show a visualization of the proximity network in a given time for the three datasets. 
Panel (e) exhibits a situation of high clusterization where the ball is in motion, panel (f) a case of high defragmentation, and panel (g) shows a moment of the game when the players are marking in the context of a free-kick.

\subsection{Temporal structure of the time series $\kappa(t)$}
\label{se:avalanches}

The evolution of $\kappa(t)$ bears essential information to describe the development of the marking dynamics.
In this section, we analyze the statistical regularities of these series.

\begin{figure*}[t!]
\centering
\includegraphics[width=0.5\textwidth]{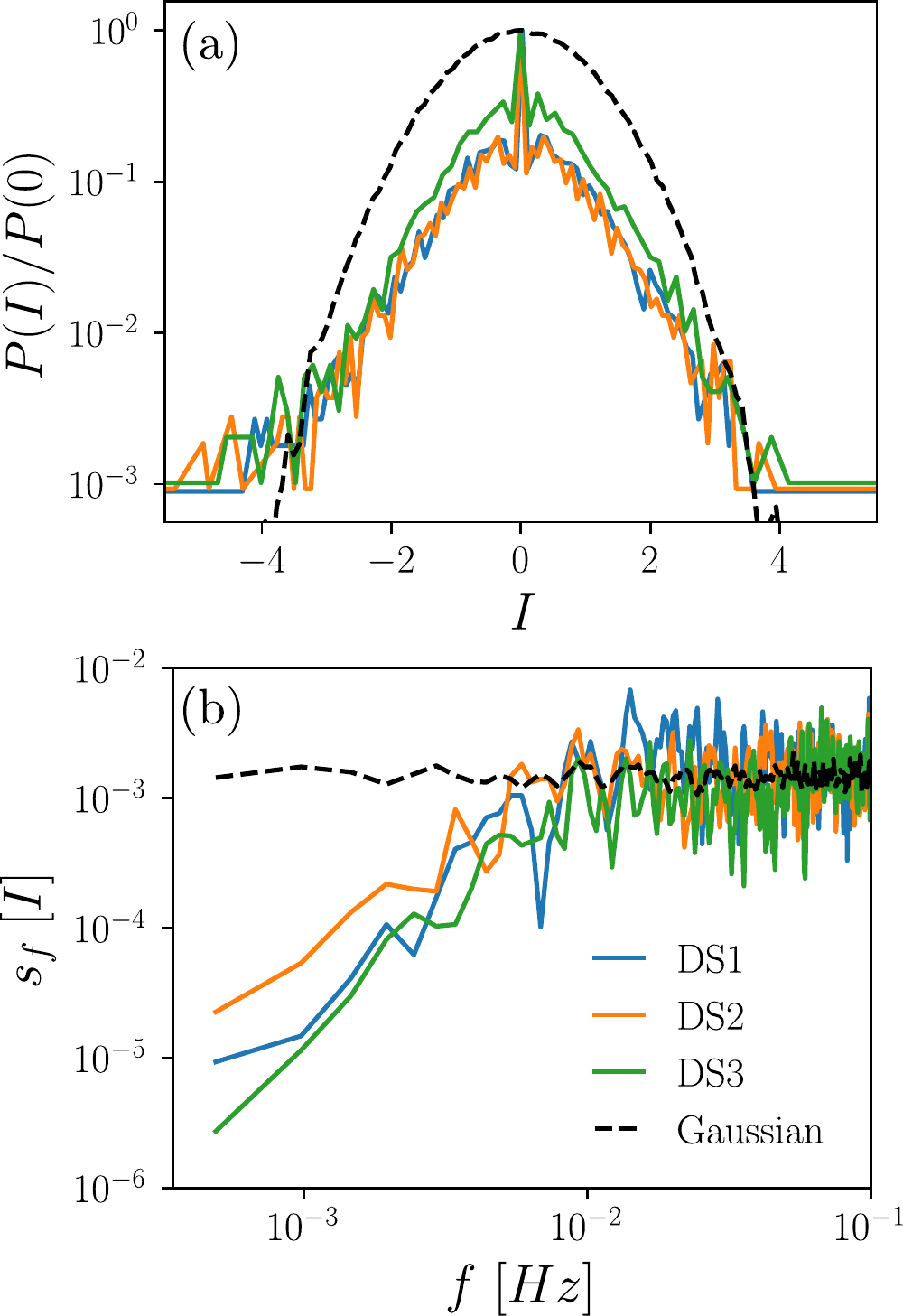}
\caption{
Statistics of the successive increments, $I$. 
We compare the results obtained for $DS1$,$DS2$, and $DS3$ with a series of Gaussian noise in both plots.
Panel (a) shows the distribution $P(I)$, and panel (b) shows the power spectrum density $s_f[I]$.
}
\label{fi:3}
\end{figure*}

Let us focus firstly on characterizing the successive increments in the series, defined as $i(t)=\kappa(t+1)-\kappa(t)$.
In Fig.~\ref{fi:3} panels (a) and (b), we show the probability density and the power spectrum density of $I(t)=i(t)/\sigma_{\kappa}$ calculated for the three datasets. In the panels, and in order to trace a comparison, we additionally show the results considering  Gaussian noise.
We observe the data deviates from the Gaussian behavior, exhibiting heavy tails in the distribution and a decay in the power spectrum density for low values of the frequency. 
Additionally, we performed a detrended fluctuation analysis (DFA) on the three series aiming to calculate the generalized Hurst exponent, $h$. We found values close to zero, which shows that the series are anti-persistent.
According to these results, the nontrivial behavior of the succession of increments reveals a complex time evolution of $\kappa$ in the three games. 

In the following, we complement these results by studying the presence of self-similar behavior in the fluctuations.
We define an event $x$ of the series $\kappa(t)$ as the consecutive points starting when $\kappa>2$ and ending when $\kappa<2$.
Note that this is equivalent to the definition of avalanches in other contexts \cite{laurson2013evolution}.
In addition, given an event $x$, we define the event lifetime $T$ as the duration of the event; and the event size $S$ as the integral under the curve.
In the following, we perform a statistical analysis of the events' lifetime and sizes for the 1050 events gathered from the time series of $\kappa(t)$ linked to the three datasets.

\begin{figure*}[t!]
\centering
\includegraphics[width=\textwidth]{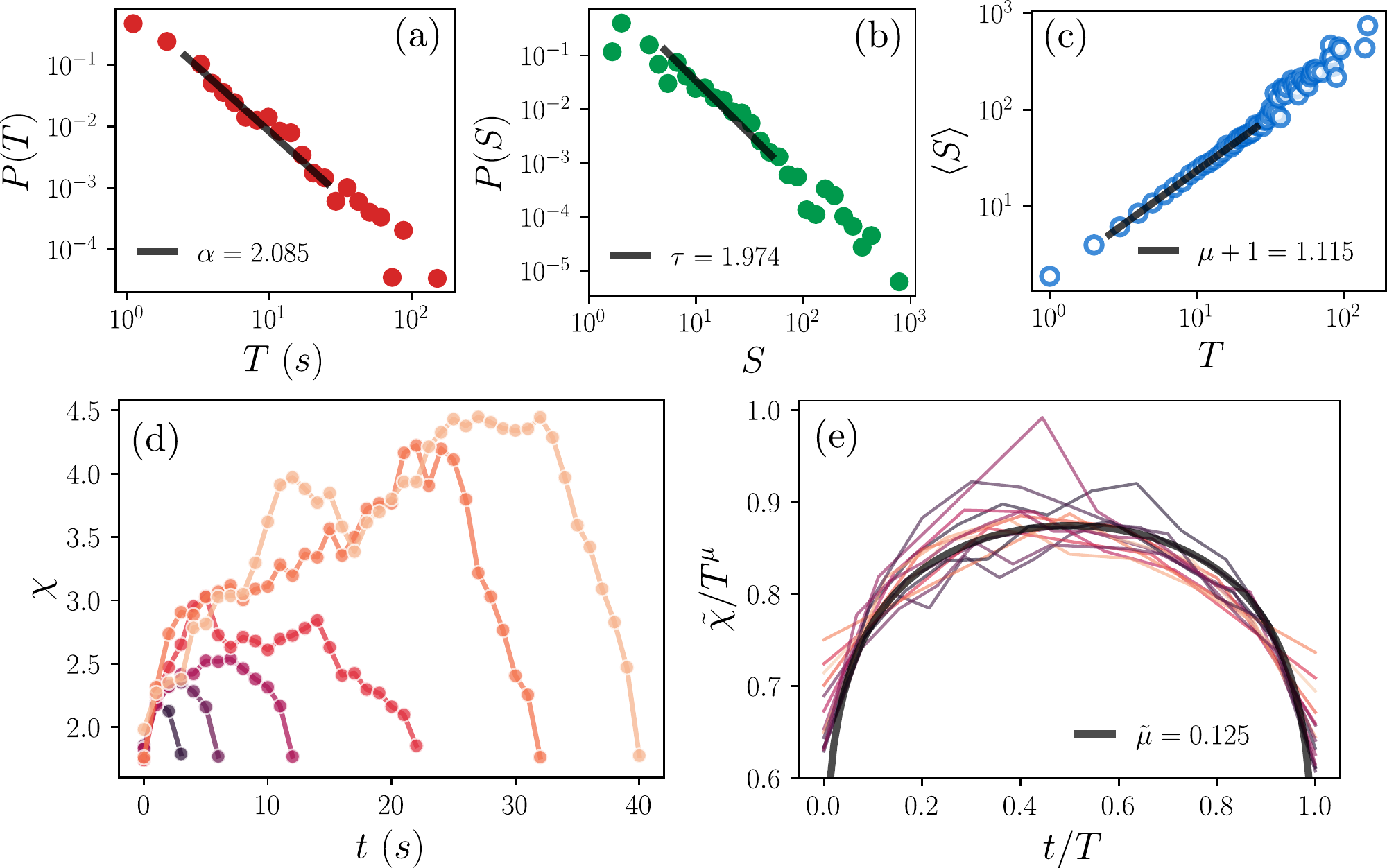}
\caption{
Self-similarity in the series of $\kappa$. 
Panels (a) and (b)show the probability distributions of avalanches lifetime and avalanches size, $P(T)$ and $P(S)$.
Panel (c) the relation between avalanches lifetime $T$ and the mean value of the size of the avalanches, $\avg{S}$.
Panel (d) several examples of avalanches with a different lifetime.
Panel (e) the collapse of the avalanches produced by the rescaling.
Black solid lines in panels (a),(b), (c), and (e) show the result of nonlinear fit in the drawn regions (see main text for further details).
}
\label{fi:4}
\end{figure*}

In Fig.~\ref{fi:4} (a), (b), and (c), we show the distribution of events' lifetime $P(T)$, the distribution of events' sizes $P(S)$, and the relation $T$~vs.~$\avg{S}$, respectively. 
We can observe a power-law behavior in the three cases. 
According to \cite{bak2013nature,chialvo2015we}, if these relations follow the universal functional forms:
\begin{equation} 
\label{eq:1}
\begin{split}
P(T) &\sim T^{-\alpha},\\
P(S) &\sim S^{-\tau},\\
\avg{S} &\sim T^{\mu+1},
\end{split}
\end{equation}
and, in addition, the following relation between the exponents holds,
\begin{equation} 
\label{eq:2}
\frac{\alpha -1}{\tau-1}= \mu+1,
\end{equation}
then, we are in the presence of a self-similar process.
To analyze this hypothesis, we perform a nonlinear fit on the empirical curves using the maximal likelihood method proposed in \cite{clauset2009power}.
We found that the empirical exponents closely fulfill Eq.~(\ref{eq:2}) for all the set of values of $T$ and $S$, particularly for the region delimited for $T \in (2, 32)~sec$ and $S \in (5,85)$, we obtain full agreement with $\alpha=2.085$, $\tau= 1.974$, and $\mu=0.115$.

Therefore, from scaling arguments \cite{baldassarri2003average}, it is expected that the average profile of an event of lifetime $T$, $\chi:=\avg{x(T,t)}$, scales as:
\begin{equation} 
\label{eq:3}
\chi= T^\mu \rho(t/T),
\end{equation}
where events of different lifetimes rescaled by the parameter $\mu$ should collapse on a single scaling function given by $\rho(t/T)$.
With this idea in mind, in Fig.~\ref{fi:4} panel (d), we show several examples of averaged events profiles with different lifetimes and, in panel (e), the collapse using Eq.~(\ref{eq:3}). 
In the latter, we normalize the profiles as $\tilde{\chi}=\chi/\chi_{MAX}$ where $\chi_{MAX}$ is the maximum  value observed  in the set of all the curves in the plot. 
With this data, we perform a nonlinear fit via the function:
\begin{equation} 
\label{eq:4}
\rho(t^\prime)=(A~t^\prime(1-t^\prime))^{\tilde{\mu}},
\end{equation}
with $t^\prime= t/T$, obtaining $A=1.37$ and $\tilde{\mu}=0.125$. 
Note that the value obtained for $\tilde{\mu}$ is consistent with the value of $\mu$ obtained from Eqs.~(\ref{eq:1}).

\subsection{Model}

\subsubsection{The equations of players' evolution}
Football dynamics can be thought of as the outcome of a particular interaction between  players. 
Teammates interact while running a tactical scheme, and opponents interact on the marking.
In this section, based on the ideas of our previous work \cite{chacoma2021stochastic}, we propose to model the players' motion in the field within the following framework: In the center of mass frame of reference, we define 
$\vec{r}_n(t)= \big(x_n(t), y_n(t)\big)^T$ and 
$\vec{v}_n(t)= \big(v_n^x(t), v_n^y(t)\big)^T$ 
as the position and velocity of player $n$ at time $t$.
We propose that every player is bounded to $(i)$ a place in the field related to their natural position in the tactical scheme of the team, $\vec{b}_n$, and $(ii)$ the other players, both teammates and opponents.
In this frame, the equation of motion for a player $n$ can be written as follow,
\begin{equation}
\centering
M_n \ddot{\vec{r}}_n = 
-\gamma_n\vec{v}_n 
+ k_{bn}(\vec{b}_n-\vec{r}_n) +
{\sum_m}^\prime k_{nm}(\vec{r}_m-\vec{r}_n),
\label{eq:eqmov}
\end{equation}
where the first term is a damping force, the second one is an ``anchor'' to the player's position, and the sum in the third term is the contributions of the interaction forces related to both teammates and opponents.
We propose different interaction constants in the horizontal and vertical axis, thus the parameters $\gamma_n$, $k_{bn}$, and $k_{nm}$, are $2D$ diagonal matrices such as, 
$\gamma_n= 
\big(\begin{smallmatrix}
  \gamma_n^x & 0\\ 
  0 & \gamma_n^y
\end{smallmatrix}\big)$,
$k_{bn}= 
\big(\begin{smallmatrix}
  k_{bn}^x & 0\\ 
  0 & k_{bn}^y
\end{smallmatrix}\big)$, and
$k_{nm}= 
\big(\begin{smallmatrix}
  k_{nm}^x & 0\\ 
  0 & k_{nm}^y
\end{smallmatrix}\big)$.
Note that these forces are not isotropic.
Moreover, since it is expected that players have similar mass, for simplicity we will consider $M_n=1$ for all the players in the field.

\subsubsection{Fitting the model's parameters}
In this section we show how to obtain the parameters $\gamma_n$, $k_{bn}$, $k_{nm}$,  and $\vec{b}_n$ by fitting Eq.~(\ref{eq:eqmov}) to the datasets. 
To perform this calculation, we considered the following:
\begin{enumerate}
    \item The velocity is calculated as $\vec{v}_n(t):=\frac{\vec{r}_n(t+\Delta t)- \vec{r}_n(t)}{\Delta t}$ ($\Delta t=1~s$).
    
    \item The discrete version of the system of first-order equations given by Eq.~(\ref{eq:eqmov}), provides the tool to estimate the states of the players at time $t+\Delta t$ by using as inputs the real states at time $t$ and the model's parameters,
    \begin{equation*}
    \begin{split}
    \vec{r}_n(t+\Delta t)^{\prime} &=
    \vec{r}_n(t)+\vec{v}_n(t)\Delta t\\
    \vec{v}_n(t+\Delta t)^{\prime} &= \vec{v}_n(t)+ \\
    & \big(-\gamma_n \vec{v}_n(t)-
    \big(k_{bn}+{\sum_m}^\prime k_{nm}\big) \vec{r}_n(t)+\\ 
    &{\sum_m}^\prime k_{nm}\vec{r}_m(t)+ k_{bn}\vec{b}_n\big) \Delta t.
    \end{split}
    \label{eq:discretesystem}
    \end{equation*}
    Where $\vec{r}_n(t+\Delta t)^{\prime}$ and $\vec{v}_n(t+\Delta t)^{\prime}$ are the model's estimations.
    
    \item Note that we are considering the definition of the velocity expressed in item $1$, $\vec{r}_n(t+\Delta t) = \vec{r}_n(t+\Delta t)^{\prime}$. 
    Then, at every step, the model's parameters are only used to predict the new velocities.
    
    \item We can choose the values of $\vec{b}_n$ such that the equilibria point of the players are their average position, $\vec{c}_n$. By doing this, we can calculate $\vec{b}_n$ using Eq.~(\ref{eq:eqmov}), the values of $\vec{c}_n$, and the other parameters.
    
    \item We define the error $\vec{\xi}_n(t) := \vec{v}_n(t+\Delta t)-\vec{v}_n(t+\Delta t)^\prime$, and fit $\gamma_n$,  $k_{bn}$, and $k_{nm}$ by minimizing the sum $\sum_t \sum_n \big|\vec{\xi}_n(t)\big|$.
\end{enumerate}

With this methodology, we obtain a unique set of parameters that control the players' motion equations and, consequently, the dynamics of the game in each dataset.

\subsubsection{Simulations}

Once performed the fit to obtain the model's parameters, we use these outcomes to simulate the coupled equation system given by Eq.~(\ref{eq:eqmov}).
To input energy into the system, we add a non-correlated Gaussian noise, such that 
$\vec{\xi}_n(t)= \vec{\sigma}_n \xi_n$, with
$\avg{\xi_n(t)}=0$,
$\avg{\xi_n(t)\xi_n(t^\prime)}= \delta (t-t^\prime)$, 
$\avg{\xi_n(t)\xi_m(t)}= 0$.
Here, for $\vec{\sigma_n}= (\sigma^x_n, \sigma^y_n)$, we use the scale of the velocity fluctuation measure in the fit. 
In this manner, the noise acts as a proxy to introduce higher-order contributions of the interaction forces into the model.

In this frame, we can write,
\begin{equation}
\begin{split}
d\vec{r}_n &= \vec{v}_n dt\\
d\vec{v}_n &= \big[
-\big(k_{nb}+{\sum_m}^\prime k_{nm}\big) \vec{r}_n
+ {\sum_m}^\prime k_{nm}\vec{r}_m -\\
&\gamma_n \vec{v}_n 
+ k_{nb}\vec{b}_n 
\big] dt + \vec{dW_n},
\end{split}
\label{eq:system2}
\end{equation}
where $d\vec{W}_n = \vec{\sigma}_n \xi_n dt$.
To solve this system of stochastic differential equations (SDE), we use the Euler--Maruyama algorithm for Ito equations.

\begin{figure*}[t!]
\centering
\includegraphics[width=\textwidth]{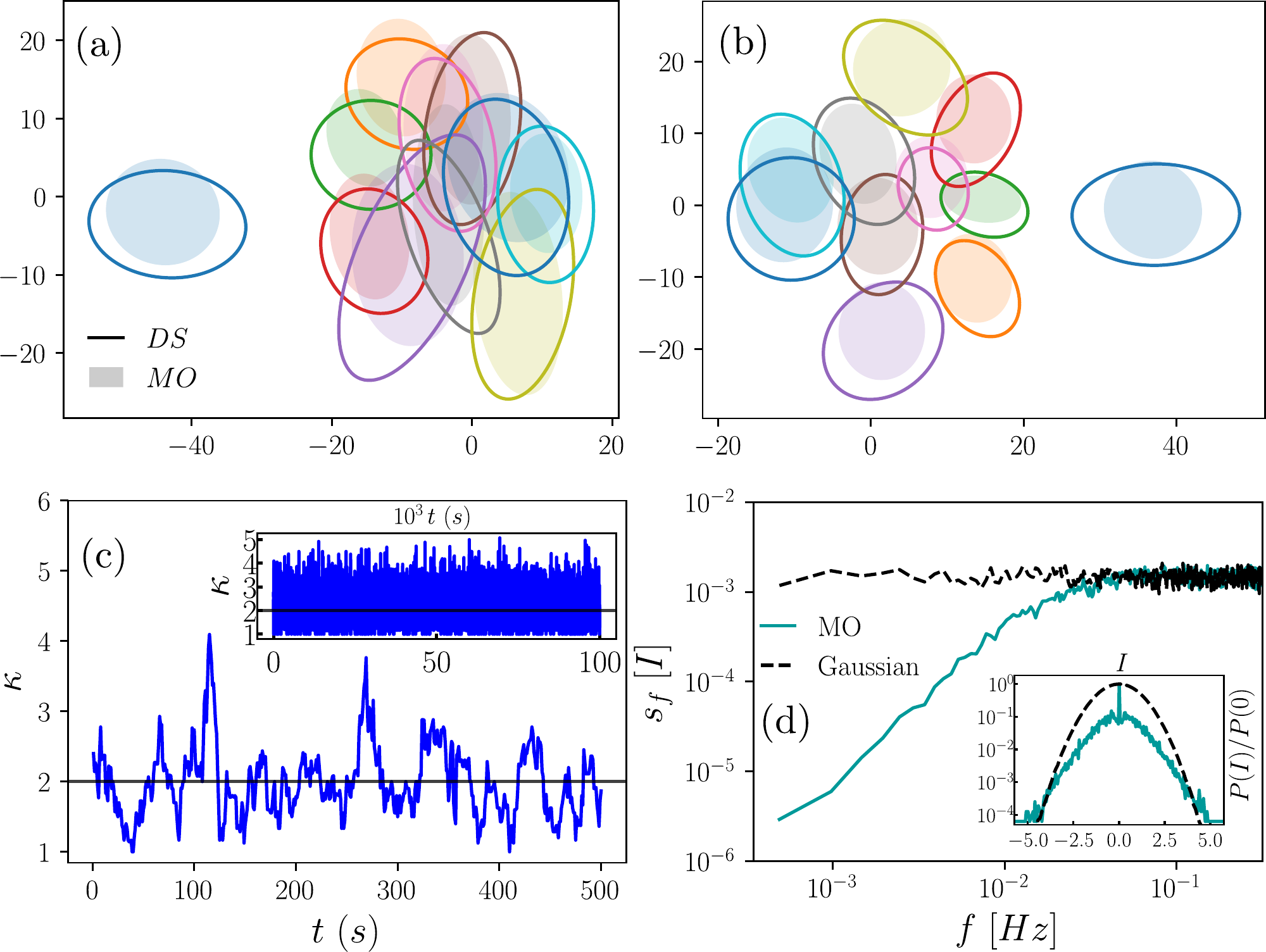}
\caption{
Outcomes of the simulations.
Panels (a) and (b) compare the empirical players' action zones of both teams with the obtained from the simulations.
Panel (c) exhibits the evolution of the heterogeneity parameter $\kappa$. In the main plot, we show a small time window of $500~sec$ where we can visualize some avalanches, whereas, in the inset, we show the entire time series.
Panel (d) shows the power spectrum density $s_f[I]$ and the distribution of I in the inset.
}
\label{fi:5}
\end{figure*}

We performed simulations using the set of parameters obtained by fitting the first half of the game recorded in $DS1$. 
We simulated a continuous game of $10^5~sec$. From this outcome, we extracted the players' trajectories to analyze.
In the following, we extend our discussion on the results.
Let us focus on Fig.~\ref{fi:5} panels (a) and (b). Here we compare the players' empirical action zones with those obtained from the simulations. 
We can see a reasonably good approximation, which indicates that the model allows us to reproduce the player's motion in the field. 
On the other hand, in panel (c), we show the time evolution of parameter $\kappa$. Notice that to calculate the proximity networks, we use $\theta=8.5$, the same value used for $DS1$. In the inset, we show the total evolution, and in the main figure, the values for the first $500~sec$. In the latter we can see the emergence of avalanches as in the empirical case.
Lastly, to analyze the temporal structure of the series, we calculate the successive increments and study, as we did in section \ref{se:avalanches}, the probability density and the power spectrum density. In panel (d), we show these results. We can see that the data deviate from the Gaussian behavior. Moreover, we performed a $DFA$ analysis to obtain the generalized hurts exponent which gave a value close to zero indicating the presence of anti-persistency. 
These results are consistent with the empirical case, indicating that the model succeeds in capturing the overall statistic of the complex evolution of the heterogeneity parameter.

\subsubsection{Analysis of avalanches in the series $\kappa(t)$ obtained from simulations}

\begin{figure*}[t!]
\centering
\includegraphics[width=\textwidth]{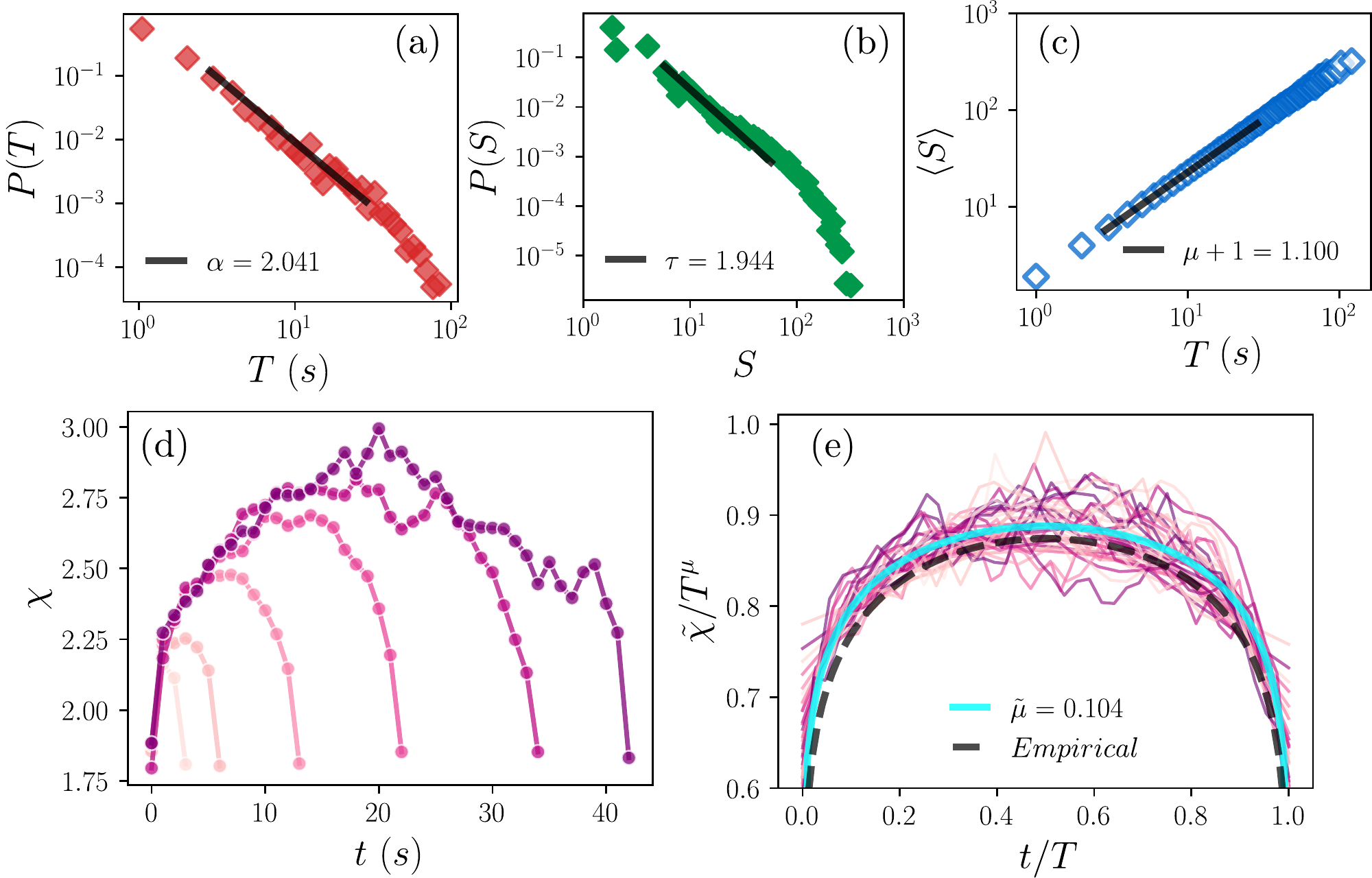}
\caption{
Self-similarity in the series of $\kappa$ obtained from simulations. 
Panels (a) and (b)show the probability distributions of avalanches lifetime and avalanches size, $P(T)$ and $P(S)$.
Panel (c) the relation between avalanches lifetime $T$ and the mean value of the size of the avalanches, $\avg{S}$.
Panel (d) several examples of avalanches with a different lifetime.
Panel (e) the collapse of the avalanches produced by the rescaling.
Black solid lines in panels (a), (b), (c), and  cyan solid lines in panel (e) show the result of nonlinear fit in the drawn regions.
Dashed line in panel (e) depicts the nonlinear fit that we have previously shown in Fig.~\ref{fi:4} (e) and is used to compare the results with the empirical case.  
}
\label{fi:6}
\end{figure*}

We repeat the analysis of self-similarity that we performed in section \ref{se:avalanches} but in this case, using the series of $\kappa$ obtained from the simulations.
In Fig.~\ref{fi:6}, panels (a), (b), and (c), we show the distributions of avalanches' lifetimes, sizes, and the relation $T$ vs. $\avg{S}$, respectively. 
In this case, we can see a cut-off in the distributions $P(T)$ and $P(S)$, particularly evident to the naked eye in the latter. 
This cut-off indicates that the model cannot generate the larger avalanches observed in the empirical case.
This is because there are particular waiting times during the game when the players pack all together in a field sector, for instance, during a corner kick or dead ball. In these moments, $\kappa$ increases shaping an avalanche that is not related to the players' motion in the field but to a particular dead time in the game. 
Our minimalist model cannot capture this phenomenon; therefore, we do not see the same tail in the distribution of avalanche sizes $P(S)$.

However, if we fit these three relations using Eqs.~(\ref{eq:1}), as we did with the empirical case, and in a similar range $T \in (2, 29)~sec$ and $S \in (5,73)$, we obtain $\alpha=2.041$, $\tau= 1.944$, and $\mu=0.1$. 
Note that these values fulfill the scaling relation expressed in Eq.~(\ref{eq:2}).
Therefore, as in the empirical case, we can write a scaling law to universally describe the avalanches' profile. In Fig.~\ref{fi:6} panels (d) and (e), we show average events of different lifetimes and the collapse of the curves into a universal form, respectively. We fit the collapse via Eq.~(\ref{eq:4}) obtaining $\tilde{\mu}=0.104$, which is consistent with the value obtained for the parameter $\mu$.
Additionally, in panel (e), we include the nonlinear fit that we have previously shown in Fig.~\ref{fi:4} (e) (see the curve in black dashed line) to exhibit the differences between the results of the model and the empirical case.

\subsubsection{The effect of the tactical system structure in the proximity networks and the evolution of $\kappa$.}

If the process $\kappa(t)$ were a Wiener process, the average shape of the fluctuations would be a semi-circle, with $\tilde{\mu}=1/2$ \cite{baldassarri2003average}.
In our case, a value $\tilde{\mu}<<1/2$ indicates anti-persistency, where the evolution of the walker suffers a restitution force that drives it to the mean value. 
%
%
%
This leads to the breakdown of the scaling laws for large times, affecting the events' lifetime $T$ and consequently the average shape of the fluctuations.
According to \cite{baldassarri2003average}, to consider this effect, we can model the evolution of $\kappa(t)$ such as,
\begin{equation}
\kappa(t+1)= \kappa(t) - \frac{1}{\tau}\; \kappa(t) + \xi(t),
\label{eq:7}
\end{equation}
where $\xi(t)$ is a random variable and the term $\frac{1}{\tau}\; \kappa(t)$ represents the effect of a parabolic well which introduces a characteristic time $\tau$ in the system.
Via the image method, one can calculate the analytical expression for the average fluctuation as a function of $t$, $T$, and $\tau$,
\begin{equation}
\chi= 
\sqrt{\frac{4\;\tau}{\pi}
\frac{(1-e^{-2t/\tau})(1-e^{-2(T-t)/\tau})}
{(1-e^{-2T/\tau})}}.
\label{eq:8}
\end{equation}

Aiming to estimate the value of $\tau$, we used Eq.~(\ref{eq:8}) to perform a nonlinear fit of the average shape of the fluctuations. From this procedure, we obtain $\tau=6.06\pm 0.07~ sec$.
Interestingly, this value is in the order of magnitude of the average ball possession time reported in \cite{chacoma2020modeling}, which is $\sim13.72~sec$. Therefore, the emergence of this time scale in the system could be related to teams moving from offensive to defensive positions or vice versa.

Note the intrinsic marking dynamic is strictly related to the anti-persistency, the players' going out from their centers to mark an opponent and returning to their action zones to cover spaces induces damped dynamics in the marking.
In other words, the constraint that the players suffer, in order to maintain the structure of the tactical system, is responsible for the anti-persistence that we see in the time series $\kappa(t)$.
Moreover, we can observe that this effect is higher in the model, where we explicitly rule the players' motion with restitution forces. It could indicate that, in the empirical case, the players actually move around with more freedom than in the model.

\section{Conclusions} 

In summarizing, we observed that the proximity network evolves following the marking dynamics, exhibiting oscillating periods of high defragmentation and high clusterization.
To characterize this phenomenon, we calculated the heterogeneity parameter and found that the system evolves in a regime similar to a transition in percolation theory.
As we previously remarked, since the system is far from the thermodynamic limit, we cannot frame our results in the theory of phase transitions. Our observations, however, evidence the emergence of complexity in the marking dynamics.
We were able to study this complex behavior by analyzing the temporal structure of the time series of $\kappa$.
We found the presence of anti-persistency  and self-similarity, which we characterized by uncovering a scaling law in the average shape of the fluctuations. 
Lastly, we proposed a model to simulate the players' motion on the field. 
From simulations, we obtained the evolution of a synthetic proximity network that we analyzed with the same methodology we used in our analysis of the empirical data.
Remarkably, the model showed a good performance in recovering the statistics of the empirical trajectories; and, consequently, the statistics of the temporal structure of the parameter $\kappa$.

The correlations observed in the proximity network associated to the marking dynamics could be related to the high level of coordination required to keep running the tactical system.
At this point, it is necessary to highlight that the marking process can not be only defined as a function of the proximity between opponents. In this sense, our study disregards several effects, hence the tactical conclusions that turn out from our analysis are limited.
For instance, the marking process can be affected by the players' positions on the field, the moment in the match, players' marking styles, etc.
Despite this limitation, our framework based on proximity networks allows us to observe that
at each game challenge, the entire team will proceed in coordination to give a response. 
They will tend to react optimally, according to the training precepts received.
Therefore, it is expected that, in similar situations, they will produce equivalent responses.
In our framework, these responses are encoded in the proximity networks as recurrent configurations and yield the memory effects we observe in the evolution of the heterogeneity parameter.

Moreover, the presence of correlations reveals the players are strongly connected.
These connections drive the team to behave flexibly and adaptable to stimuli, something crucial for the development of the game.
We can compare this "state of alert" of the teams with what occurs with a bird flocks or a fish shoals, in which connections among the individual make the group stronger to avoid predators \cite{albano1996self,cavagna2010scale,hattori1999self,juanico2007dissipative}.
The difference between these cases and the dynamics of a football team relays on the cognition capabilities required to achieve this level of organization among the group's individuals.

The emergence of complexity in the game of football is somewhat similar to that observed in a living system. 
In these systems, when is unbalanced the delicate equilibrium between inhibition and promotion, cooperation and competition, something abnormal occurs. 
This effect is observed, for example, in the appearance of cancer cells \cite{SEDIVY1999271}, in diseases of the nervous system \cite{moustafa2016complexity}, in diseased mitochondria \cite{zamponi2018mitochondrial}, etc. 
When the complexity of the system is lacking, its functioning is severely damaged.
Analogously, in the case of football dynamics, the lack of complexity would be related to low level played games. 
Therefore, our framework provides a tool that can help to detect a lack of performance in the teams.

Finally, we point out that other tactical-oriented types of analysis can be performed via the study of the temporal proximity network.
For instance, the value of $\theta$ that brings the marking system to the critical threshold could indicate the type of marking: A low value can correspond to man-to-man marking, whereas a high value to a zone or a hybrid system. 
In the same line, it is possible to study the players' performance by characterizing recurrent configurations in the networks and the formation of small communities. 
In this regard, we let the door open to further studies on this topic.

\section*{Acknowledgement}

This work was partially supported by CONICET under Grant number PIP 112 20200 101100; FonCyT under Grant number PICT-2017-0973; and SeCyT-UNC (Argentina).

\clearpage
\newpage

\end{document}